\documentclass{spie}%
\usepackage{amsfonts}
\usepackage{amsmath}
\usepackage{amssymb}
\usepackage{graphicx}%
\begin{document}

\title{Quantum Hidden Subgroup Algorithms: The Devil Is in the Details}
\author{Samuel J. Lomonaco, Jr.\supit{a} and Louis H. Kauffman\supit{b}
\skiplinehalf\supit{a}Department of Computer Science and Electrical
Engineering, University of Maryland
\and \ Baltimore County, 1000 Hilltop Circle, Baltimore, Maryland 21250,USA\\\supit{b}Department of Mathematics, Statistics, and Computer Science,
University of Illinois \\at Chicago, 851 South Morgan\ Street, Chicago, Illinois 60607-7045,USA }
\authorinfo{Further author information: S.J.L., Jr. E-mail: Lomonaco@umbc.edu: L.H.K.
E-mail: kauffman@uic.edu}
\maketitle

\begin{abstract}
We conjecture that one of the main obstacles to creating new non-abelian
quantum hidden subgroup algorithms is the correct choice of a transversal.

\end{abstract}

\section{Introduction}

\bigskip

New quantum algorithms are crucially needed to demonstrate that future quantum
computers will be more than highly specialized machines, but instead highly
versatile general purpose devices. \ Unfortunately, the current choice of
quantum algorithms available for future quantum computing devices is
surprisingly meager. With some minor (but important) exceptions, there are
only three available classes of quantum algorithms, namely:

\begin{itemize}
\item[\ ] 1) Quantum hidden subgroup (QHS) algorithms, i.e., Shor/Simon-like algorithms,

\item[\ ] 2) Amplitude amplification algorithms, i.e., Grover-like algorithms, and

\item[\ ] 3) Quantum algorithms that simulate quantum systems on a quantum computer.
\end{itemize}

\bigskip

In this paper, we focus on the development of new non-abelian QHS algorithms
because of their tantalizing promise of exponential speedup over existing
classical algorithms. \ For example, should there be a major breakthrough in
non-abelian QHS algorithm development, then there is the enticing possibility
that it might be possible to develop a polynomial-time QHS algorithm for the
graph isomorphism problem.

\bigskip

In spite of the efforts of many researchers, the number of new non-abelian QHS
algorithms is very small and scattered. Why is it so difficult to generalize
QHS algorithms to non-abelian groups?

Before we can even guess at an answer to this question, we first need to
define what we mean by a quantum hidden subgroup algorithm.

\bigskip

\section{The quantum hidden subgroup algorithm paradigm}

\bigskip

We begin by first defining the classical hidden subgroup problem:

\bigskip

\begin{itemize}
\item[\ ] \textsc{Hidden Subgroup Problem.} Given a group $G$ and a map
$f:G\longrightarrow H$ (not necessarily a morphism) of $G$ into a group $H$,
is it possible to find an invariant subgroup $K$ of $G$ such that the map $f$
can be factored into the composition $f=\iota\circ\gamma$ of two maps
$\gamma:G\longrightarrow G/K$ and $\iota:G/K\longrightarrow H$, where $\gamma$
denotes the natural epimorphism, and where $\iota$ denotes an injection. The
subgroup $K$ is called a \textit{hidden subgroup}.
\end{itemize}

\bigskip

The quantum analog of the classical hidden problem is defined as follows:

\bigskip

\begin{itemize}
\item[\ ] \textsc{Generic Quantum Hidden Subgroup Problem.} Let $\mathcal{H}%
_{G}$ and $\mathcal{H}_{H}$ be Hilbert spaces with respective orthonormal
bases%
\[
\{\left\vert g\right\rangle :|g\in G\}\text{ and }\{\left\vert h\right\rangle
:h\in H\}\text{ .}%
\]
Assume that $f:G\longrightarrow H$ is given as a unitary transformation%
\[%
\begin{array}
[c]{rrl}%
U_{f}:\mathcal{H}_{G}\otimes\mathcal{H}_{H} & \longrightarrow & \mathcal{H}%
_{G}\otimes\mathcal{H}_{H}\\
\left\vert g\right\rangle \left\vert h\right\rangle \quad\  & \longmapsto &
\left\vert g\right\rangle \left\vert f(g)h^{-1}\right\rangle
\end{array}
\]
Determine the hidden subgroup of $f$ by making as few queries as possible to
the blackbox $U_{f}$.
\end{itemize}

\bigskip

We can now in turn define what we mean by a QHS algorithm that solves the
above QHS problem:

\bigskip

The \textsc{generic quantum hidden subgroup (QHS) algorithm} is essentially
(ignoring many subtleties) the following:

\begin{itemize}
\item \textsc{Step O.} Initialize two registers $|$\textsc{Left-Reg}$>$ in
$\mathcal{H}_{G}$ and $|$\textsc{Right-Reg}$>$ in $\mathcal{H}_{H}$ to produce
the initial state $\left\vert \psi_{0}\right\rangle =\left\vert 0\right\rangle
\left\vert 1\right\rangle $, assuming additive notation for $G$, and
multiplicative notation for $H$.

\item \textsc{Step 1.} Apply the Fourier transform $\mathcal{F}_{G}$ to the
left register.

\item \textsc{Step 2.} Apply the unitary transformation $U_{f}$ . (As
mentioned by Shor \cite{Shor1}, there is no need to measure the right register.)

\item \textsc{Step 3.} Again apply the Fourier transform $\mathcal{F}_{G}$ to
the left register.

\item \textsc{Step 4.} Measure the left register.

\item \textsc{Step 5.} Repeat Steps 0 through 4 until enough measurements have
been made to determine the hidden subgroup.
\end{itemize}

\bigskip

Both Jozsa \cite{Jozsa3} and Kitaev \cite{Kitaev2} have noted that Simon's
quantum algorithm and Shor's quantum factoring algorithms can be viewed as
instances of the same generic quantum hidden subgroup algorithm for abelian
groups. Simon's algorithm determines a hidden subgroup of the direct sum of
cyclic groups of order 2. Shor's algorithm determines a hidden subgroup of the
infinite cyclic group.

\bigskip

But can the generic abelian hidden subgroup algorithm be extended to a larger
class of groups? If it could be extended to a large enough class of
non-abelian groups, then there is a strong possibility that a polynomial-time
quantum algorithm for the graph isomorphism problem could be found in this
way. (For an in depth study of the graph isomorphism problem, see, for
example, Hoffman\cite{Hoffmann1}. For applications, see, for example,
Tarjan\cite{Tarjan1}. For a discussion as to how to extend the quantum hidden
subgroup problem to non-abelian groups, see for example
Lomonaco\cite{Lomonaco8}.)

\bigskip

But once again, we ask:%
\[
\text{Why has it been so difficult to create new non-abelian QHS algorithms?}%
\]

\ We can not give a complete answer to this question. \ But we can give some
clues as to why non-abelian QHS algorithm development has been so difficult.

\bigskip

\section{\textquotedblleft The devil is in the details.\textquotedblright}

\bigskip

\textquotedblleft The devil is in the details.\textquotedblright\ By that we
mean the following:

\bigskip

A number of subtleties were ignored in the above description of the generic
QHS algorithm. These subtleties become absolutely crucial when one tries to
extend the generic QHS algorithm to a larger class of non-abelian groups.

\bigskip

There is simply not enough space to discuss all the important subtleties in
regard to the generic QHS algorithm. So we will focus on only one, which the
authors believe has been ignored by many in the research community.

\bigskip

Contrary to conventional wisdom, Simon's and Shor's algorithms are far from
the same generic QHS algorithm. To say so is a gross oversimplification. There
are crucial differences. Shor's algorithm, unlike Simon's, outputs group
characters which are approximations to the group characters actually sought.

\bigskip

More specifically, to find a hidden subgroup of $f:G\longrightarrow H$, Shor's
algorithm selects a \textquotedblleft large\textquotedblright\ epimorphic
image $\nu:G\longrightarrow Q$ for \textquotedblleft
approximating\textquotedblright\ $f$. Crucially, Shor's algorithm also selects
a \textquotedblleft correct\textquotedblright\ transversal map $\tau
:Q\longrightarrow G$ , where a transversal $\tau:Q\longrightarrow G$ is an
injection such that $\nu\circ\tau=id_{Q}$, where $id_{Q}$ denotes the identity
map on $Q$.

\bigskip

Once the transversal $\tau$ has been selected, we can define an approximation
$\widetilde{f}$ to $f$ by $\widetilde{f}=f\circ\tau$. We can now use the
Fourier transform of $\widetilde{f}$ associated with the group $Q$ to find the
hidden subgroup of $\widetilde{f}$. Thus, Shor's algorithm outputs random
group characters of $Q$ which are approximations to the group characters of
$H$.

\bigskip

We emphasize one key point here:
\[
\text{The selection of the transversal }\tau:Q\longrightarrow G\text{ \ .\ }%
\]
On the one hand, a good choice will produce an efficient algorithm. On the
other hand, a bad choice will lead to an inefficient algorithm.
\ Unfortunately, Shor's original algorithm gives no hint as to how to choose
the transversal $\tau:Q\longrightarrow G$ generically. This, of course, is
only one of the difficulties encountered in trying to generalize Shor's
algorithm to non-abelian groups. \ But it appears to be a crucial one that
has, for the most part, been ignored. \ The significance of the correct choice
of a transversal is made transparent by Lomonaco and Kauffman\cite{Lomonaco5},
where the correct transversal for the abelian QHS algorithm is found to be
what is called in their paper a Shor transversal.

\bigskip

Another obstacle to the development of new non-abelian QHS algorithms is the
time complexity of the non-abelian Fourier transform.

\bigskip

\section{The non-abelian Fourier transform}

\bigskip

The Fourier transform on non-abelian groups is defined as follows:

\bigskip

Let $G$ be a finite non-abelian group, and let $\pi^{(1)}$, $\pi^{(2)}$,
$\ldots$\ , $\pi^{(k)}$ be a complete set of distinct irreducible
representations of the group $G$.

\bigskip

Each irreducible representation is a morphism $\pi^{(i)}:G\longrightarrow
Aut\left(  W_{i}\right)  $ from $G$ to the group of automophisms of the
representation space $W_{i}$ . Let $\pi^{(i)\#}:G\longrightarrow Aut\left(
W_{i}^{\ast}\right)  $, denote the corresponding contragradient
representation, where $W_{i}^{\ast}$ is the dual $W_{i}$ . Extend the
representation $\pi^{(i)\#}:G\longrightarrow Aut\left(  W_{i}^{\ast}\right)  $
to the group ring, i.e., to $\pi^{(i)\#}:\mathbb{C}G\longrightarrow End\left(
W_{i}^{\ast}\right)  $, where $\mathbb{C}G$ denotes the group ring of $G$ over
the complex numbers $\mathbb{C}$, and where $End\left(  W_{i}^{\ast}\right)  $
denotes the ring of endomorphisms of $W_{i}^{\ast}$. Then the
\textit{non-abelian Fourier transform} $\mathcal{F}_{G}$ on the group $G$ is
defined as%
\[
\mathcal{F}_{G}=%
%TCIMACRO{\dbigoplus \limits_{i=1}^{k}}%
%BeginExpansion
{\displaystyle\bigoplus\limits_{i=1}^{k}}
%EndExpansion
\pi^{(i)\#}:\mathbb{C}G\longrightarrow%
%TCIMACRO{\dbigoplus \limits_{i=1}^{k}}%
%BeginExpansion
{\displaystyle\bigoplus\limits_{i=1}^{k}}
%EndExpansion
End\left(  W_{i}^{\ast}\right)  \text{ \ ,}%
\]
where $%
%TCIMACRO{\dbigoplus }%
%BeginExpansion
{\displaystyle\bigoplus}
%EndExpansion
$ denotes the direct sum.

\bigskip

The above Fourier transform can be expressed more explicitly, but less
transparently, in terms of matrices.

\bigskip

The algorithmic time complexity of the non-abelian Fourier transform depends
to a large extent on the basis chosen to express the Fourier transform in
terms of matrices. \ But this complexity will depend on which transversal
$\tau:Q\longrightarrow G$ is chosen for the non-abelian QHS algorithm! \ 

\bigskip

\section{Conclusion}

\bigskip

In this paper, we conjecture that one of the main obstacles to creating new
non-abelian QHS algorithms for the hidden subgroup problem $f:G\longrightarrow
H$ is the correct choice for the transversal $\tau:Q\longrightarrow G$.

\bigskip

\section{Acknowledgements}

\bigskip

This work was supported by the Defense Advanced Research Projects Agency
(DARPA) and Air Force Research Laboratory, Air Force Materiel Command, USAF,
under agreement F30602-01-2-05022. Some of this effort was also sponsored by
the National Institute for Standards and Technology (NIST). The U.S.
Government is authorized to reproduce and distribute reprints for Government
purposes notwithstanding any copyright annotations thereon. The views and
conclusions contained herein are those of the authors and should not be
interpreted as necessarily representing the official policies or endorsements,
either expressed or implied, of the Defense Advanced Research Projects Agency,
the Air Force Research Laboratory, or the U.S. Government. (Copyright 2004.)
\ The first author would also like to thank the Mathematical Sciences Research
Institute (MSRI) at Berkeley, California for its support of this work.

\end{document}